\documentclass{iaus}
\usepackage{graphicx}

\title[Alpha-Enhanced SBF] 
{The Effect Of Alpha-Element Enhancement On Surface Brightness Fluctuation  
Magnitudes And Broad-Band Colors}

\author[H.-c. Lee et al.]   
{Hyun-chul Lee$^1$, Guy Worthey$^1$, \and John P. Blakeslee$^1$}

\affiliation{$^1$Washington State University \break email: hclee@wsu.edu\\[\affilskip]}

\pubyear{2007}
\volume{241}  
\pagerange{1--2}
\date{?? and in revised form ??}
\jname{Stellar Populations as Building Blocks of Galaxies}
\editors{Peletier R.F. \& Vazdekis A., eds.}
\begin{document}

\maketitle

\begin{abstract}
We present the effects due to alpha-element enhancement on surface
brightness fluctuation magnitudes and broad-band colors in order to
investigate the calibration zero-point. We study these effects at ages
covering 0.1 to 13 Gyr and metallicities of Z=0.0003 to 0.04 using the
Teramo group's isochrones, BaSTI. Our preliminary results indicate that
alpha-enhanced models are needed to match massive red galaxies while
solar-scaled models are fine for dwarf blue galaxies
to reproduce their independently estimated ages and metallicities.
\end{abstract}

\firstsection 
\section{Introduction}
The surface brightness fluctuation (SBF) magnitudes and 
colors are widely used as one of the most powerful distance indicators 
as well as useful tools for probing stellar populations.  
To our continued consternation, however, only solar-scaled SBF 
models have been calculated in the past.  

\section{Our Preliminary Results}
Here we report, for the first time, alpha-element enhanced 
SBF models and compare them with observations. For those purposes, 
we have employed the Teramo Isochrones (Pietrinferni et al. 2004, 2006; 
Cordier et al. 2007; http://www.te.astro.it/ 
BASTI ) in this study.  In general, alpha-element enhanced SBF magnitudes 
are brighter and their broad-band colors are redder where the 
alpha-element enhancement is defined at a fixed total 
metallicity (cf., http:// www.astro.rug.nl/$\sim$sctrager/FTSPM/
Lee$\_$alpha.pdf ).  
Our results also illustrate the dramatic impact of convective core 
overshooting on the predicted broad-band colors for models 
with ages near 1 Gyr (see also http://astro.wsu.edu/hclee/ 
hclee$\_$208$\_$AAS$\_$10$\_$07.pdf ). 

Figure 1 shows our preliminary results (aes: alpha-enhanced standard, 
sss: solar-scaled standard, where `standard' means no convective 
core overshooting) in the V $-$ I  vs. SBF I-band Magnitude.  
Two observational fiducial lines (thick straight lines) are overlaid 
with our theoretical models.  The blue line is
SBF I-band Mag = $-$2.13 + 2.44 $\times$ [(V $-$ I) $-$ 1.00]
from Mieske, Hilker, \& Infante (2006) 
for dwarf blue galaxies, while the red line is
SBF I-band Mag = $-$1.68 + 4.5 $\times$ [(V $-$ I) $-$ 1.15]
from Mei et al. (2005, originally calibrated at Blakeslee et al. 2002) 
for massive red galaxies.
From the comparisons, it seems that our preliminary results indicate that  
alpha-enhanced models are needed to match massive red galaxies while
solar-scaled models are fine for dwarf blue galaxies
to reproduce their independently estimated ages and metallicities.
The full investigation will be presented at Lee, Worthey, \& Blakeslee (2007).

\begin{figure}
 \includegraphics[width=.99\textwidth]{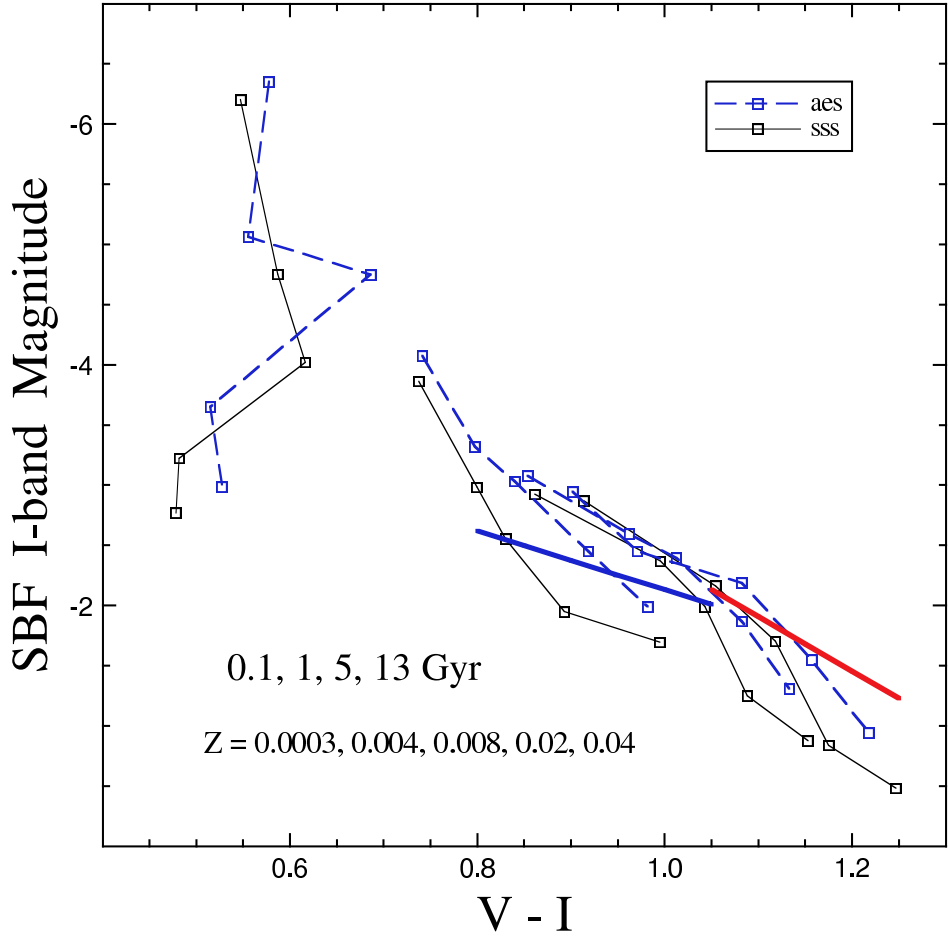}
  \caption{V $-$ I vs. SBF I-band Magnitude at 0.1, 1, 5, and 13 Gyr 
(left to right). Solid lines with squares are solar-scaled models  
while dashed lines with squares are alpha-enhanced models. 
At given ages, SBF I-band Magnitudes become fainter and 
V $-$ I colors become redder, in general, with increasing metallicity. 
Two observational fiducial lines are compared with our theoretical models 
(see text).}
\end{figure}

\begin{acknowledgments}
Support for this work was provided by the NSF through grant AST-0307487,
the New Standard Stellar Population Models (NSSPM) project. H.-c. Lee also
thanks the organizers for their wonderful hospitality and financial support.
\end{acknowledgments}

\end{document}